\documentclass[epj]{svjour}

\usepackage{amsmath}
\usepackage{bbm}
\usepackage{graphicx}
\usepackage{fancyhdr}
\usepackage{axodraw}
\usepackage{colordvi}
\def\makeheadbox{{%  remove EPJC header
 }}
\def\mytitle{My title} 
\def\myauthors{My name}  
\def\mytype{My type of session}
\def\mysession{My session}
\def\mytitle{Local $\mathsf{SU(5)}$ Unification in 6D from the Heterotic String} %Put your title here!
\def\myauthors{Christoph L\"{u}deling}    %Put your name here!
\def\mytype{Contributed Talk}    
\def\mysession{Theoretical Models}

\newcommand{\SO}[1]{\ensuremath{\mathsf{SO(#1)}}}
\newcommand{\SU}[1]{\ensuremath{\mathsf{SU(#1)}}}
\newcommand{\U}[1]{\ensuremath{\mathsf{U(#1)}}}
\newcommand{\E}[1]{\ensuremath{\mathsf{E_#1}}}

\begin{document}
\title{Local \boldmath$\SU5$ Unification in 6D from the Heterotic String}
\author{Christoph L\"{u}deling
\thanks{\emph{Email:} {\ttfamily c.luedeling@thphys.uni-heidelberg.de}}}%
% \and
% Second author\inst{2}% etc
%}                     % Do not remove
\institute{Institut f\"{u}r theoretische Physik, Universit\"{a}t Heidelberg, Philosophenweg 16 und
  19, 69120 Heidelberg, Germany}
\date{}
\abstract{
We present a six-dimensional $T^2/\mathbbm{Z}_2$ orbifold model which arises as an intermediate step
in the compactification of the heterotic string to the MSSM. The orbifold contains two pairs of inequivalent
fixed points, with unbroken local gauge groups $\SU5$ and $\SU2\times \SU4$, respectively, the
intersection of which gives the standard model gauge group. All bulk and brane anomalies are
cancelled by the Green--Schwarz mechanism. At each $\SU5$ fixed point, there is a localised 
$\bar{\boldsymbol{5}}\oplus\boldsymbol{10}$ standard model generation, while the third generation
and the Higgs fields come in split bulk multiplets due to the breaking of $\SU5$ at the other fixed point.
 % end of PACS codes
} %end of abstract
\maketitle
%DO NOT REMOVE THIS LINE
%

\section{Introduction}
\label{intro}
An attractive route for physics beyond the standard model is provided by the idea of grand
unification where the standard model group $G_\text{SM}=\SU3\times \SU3\times \U1$ is
realised as a subgroup of a larger semisimple group $G_\text{GUT}$. This idea is supported
by the observed unification of gauge couplings in the (supersymmetric) standard model at the GUT
scale $M_\text{GUT}\simeq 10^{16}$~GeV.  In the simplest case,
$G_\text{GUT}= \SU5$, and one standard model family fits into a
$\bar{\boldsymbol{5}}\oplus\boldsymbol{10}$ representation of $\SU5$. Extending this approach via
larger groups such as $\mathsf{SO(10)}$, one  arrives at $\E8$, which is realised in the
heterotic string. However, there are drawbacks to this na\"{i}ve picture of four-dimensional grand
unification. For example, large Higgs representations are required to break the GUT group to the
standard model. Furthermore, the standard model Higgs doublet comes in a $\boldsymbol{5}$ of $\SU5$,
together with a colour triplet which needs to get a mass of the order of the GUT scale to avoid
proton decay while the doublet stays light.  Phenomenologically, the unification of matter in larger multiplets
predicts a unification of Yukawa couplings at the GUT scale which is not observed.

These problems can be addressed in higher-di\-men\-si\-o\-nal orbifold GUTs, where symmetry breaking and
doublet--triplet--splitting can be achieved via projection conditions. A promising possibility for
such GUTs the heterotic string, which includes an $\E8\times\E8$ gauge symmetry and allows for
comparatively simple compactifications on orbifolds with realistic matter content and gauge groups
in the four-dimensional limit. At the same time, it guarantees the absence of anomalies and provides
a UV completion for the higher-dimensional effective field theory.

\section{\boldmath$T^2/\mathbbm{Z}_2$\unboldmath Orbifold Model}\label{sec:1}
Our model \cite{bls07} is based on the compactification of the heterotic string on the orbifold
$T^6/\mathbbm{Z}_{6-\text{II}}$, where the $T^6$ is specified by the Lie algebra lattice of
$\mathsf{G}_2 \times \SU3\times \SO4$ \cite{krz04}. This model is known to give the MSSM in the
four-dimensional limit\cite{bhx05,bhx06}. We take the limit in which the $\mathsf{G}_2
\times \SU3$ tori become small (of
$\mathcal{O}\!\left(M_\text{string}^{-1}\right)$) while the $\SO4$
torus stays larger,
$\mathcal{O}\!\left(M_\text{string}^{-1}\right)$. Hence, we end up
with an effective six-dimensional theory on $T^2/\mathbbm{Z}_2$. This
orbifold has four fixed points, but due to one Wilson line, they come
in two inequivalent pairs.

The bulk gauge group is 
\begin{align}
  \begin{split}
    G_\text{bulk}&= \SU6\times \U1^3 \\
    &\quad\mspace{70mu}\times \left[\SU3 \times \SO8 \times \U1^2\right]\,.
  \end{split}
\end{align}
Brackets denote subgroups of the second $\E8$. Bulk matter comprises the untwisted sector and the
sector twisted by the $\mathbbm{Z}_3$ subtwist of $\mathbbm{Z}_{6-\text{II}}$, whose fields are
localised at fixed points in the $\mathsf{G}_2$ and $\SU3$ tori. The untwisted sector contains the
supergravity and dilaton multiplet $\left(G_{MN}, B_{MN}, \Phi, \Psi_M,\chi\right)$, the vector
multiplets corresponding to $G_\text{bulk}$ and hypermultiplets transforming as
\begin{align}
  \begin{split}
    &\left(\boldsymbol{20};1,1\right) + \left(1;1,\boldsymbol{8}\right)+ \left(1;1,\boldsymbol{8}_s\right) \\
    &\mspace{120mu} +
    \left(1;1,\boldsymbol{8}_c\right) +4 \times \left(1;1,1\right)
  \end{split}
\end{align}
under the non-Abelian group factors. (We suppress $\U1$ charges here. The complete list of all
states and charges is given in \cite{bls07}.) For the untwisted sector, we note that at each fixed point
in the $\SU3$ torus, there is a local $\SO{14}\times \U1 \times\left[\SO{14}\times \U1 \right]$
gauge group (differently embedded into $\E8\times \E8$ each time due to a Wilson line), and  localised hypermultiplets
transforming as $\left(\boldsymbol{14};1\right)$ and $\left(1;\boldsymbol{14}\right)$. With respect
to the six-dimensional gauge group, they split into 
\begin{subequations}
\begin{align}\label{eq:t2t4}
  \left(\boldsymbol{14};1\right) &= \left(\boldsymbol{6};1,1\right) +
  \left(\bar{\boldsymbol{6}};1,1\right) +2 \times \left(1;1,1\right)\,,\\
  \left(1;\boldsymbol{14}\right) &= \left(1;\boldsymbol{3},1\right) +
  \left(1;\bar{\boldsymbol{3}},1\right) + \left(1;1,\hat{\boldsymbol{8}} \right)\,.
\end{align}
\end{subequations}
Here $\hat{\boldsymbol{8}}$ refers to $\boldsymbol{8}$, $\boldsymbol{8}_s$ and $\boldsymbol{8}_c$,
at the three fixed points, respectively. Furthermore, at each fixed point there are two non-Abelian
singlet oscillator hypermultiplets. Note that the $\mathbbm{Z}_3$  twisted sector comes in three copies due to
the three fixed points in the $\mathsf{G}_2$ torus.

The fixed points in the $\SO4$ torus are labelled by two numbers, \mbox{$n_2,n_2'=0,1$}. Points with the same value of $n_2$ 
are equivalent. At the fixed points, the $\mathbbm{Z}_2$ projection conditions break the gauge group
as indicated in Table~\ref{tab:gaugegroups}.

\begin{table}
\caption{Local gauge groups and their intersection. The standard model is part of $\SU5$ at
  $n_2=0$.}
\label{tab:gaugegroups}       % Give a unique label
\begin{tabular}{cc}
\hline\noalign{\smallskip}
$n_2$ &  Gauge group  \\
\noalign{\smallskip}\hline\noalign{\smallskip}
0 & $\SU5\times \U1^4\times \left[\SU3\times \SO8 \times \U1^2\right]$ \\
1 & $\SU4\times \SU2\times \U1^4\times \left[\SU4'\times \SU2'\times \U1^4\right] $\\
$\cap$ & $ \SU3\times \SU2\times \U1^5\times \left[\SU4'\times \SU2'\times \U1^4\right] $\\
\noalign{\smallskip}\hline
\end{tabular}
% Or use
\vspace*{1cm}  % with the correct table height
\end{table}

At the fixed points, there are chiral multiplets from the $\mathbbm{Z}_2$--twisted sector. At
$n_2=0$ they include one standard model generation as $\bar{\boldsymbol{5}} \oplus \boldsymbol{10}$
of $\SU5$, plus further $\SU5$ singlets (and no exotics). At $n_2=1$, there are only exotic states
and singlets.
%\begin{align}
%  \left(\bar{\boldsymbol{5}};1,1\right)+  \left(\boldsymbol{10};1,1\right) + 2\times
%  \left(1;\boldsymbol{3},1\right)+  2\times \left(\bar{1;\boldsymbol{3}},1\right)+
%  \left(1;1,\boldsymbol{8}_c\right) + 9\times \left(1;1,1\right) 
%\end{align}

\section{Anomalies and FI Terms}
In the anisotropic limit of the $\SO4$ torus being much larger than the other four internal
dimensions, the effective theory is a six-dimensional supergravity on $T^2/\mathbbm{Z}_2$. Such a theory
faces stringent constraints from the absence of anomalies. In particular, the condition that all
anomalies can be cancelled by the Green--Schwarz mechanism requires that the anomaly polynomials in
the bulk and at the fixed points, $I_8^\text{bulk}$ and $I_6^{n_2}$, are reducible,
\begin{align}
  I_8^\text{bulk} &= X_4 Y_4\,, & I_6^{n_2}&=\left. X_4\right|_{n_2} Y_2^{n_2}\,.
\end{align}
Here $X_4$ is fixed by the variation of the 2-form field
$B_{MN}$. Note that there are two independent fixed point anomaly
polynomials since the spectrum does not depend on $n_2'$. These equations represent
$\mathcal{O}\!\left(400\right)$ conditions on 33 free parameters in $Y_4$ and $Y_2^{n_2}$, thus the
system is strongly overconstrained. Luckily, anomaly freedom is guaranteed by string theory and
modular invariance conditions on the twist vectors and Wilson lines used to define the
model. Nevertheless, the calculation of the anomaly polynomials is worthwhile not only as a check of
the spectrum: It allows to determine the localised  two-forms $Y_2^{n_2}\sim F_2^{n_2\text{an}}$,
which are the field strengths of the local anomalous $\U1$'s. These in turn induce localised FI
terms 
\begin{align}
  \xi_0 &=2 \frac{g M_\text{P}^2}{384\pi^2}\,, &\xi_1& = \frac{g M_\text{P}^2}{384\pi^2}\,.
\end{align}
These FI terms need to be cancelled for a supersymmetric vacuum configuration and thus can stabilise
flat directions of the potential, albeit at a very high scale. Furthermore, they can induce
nontrivial profiles of bulk vevs in the internal dimensions\cite{lnz03}. However, a detailed analysis of these
issues is beyond this work.

\section{Decoupling and Local GUT}
\subsection{Decoupling the Exotics}
Bulk matter fields come in hypermultiplets, which, from the point of view of
$\mathcal{N}=1$ supersymmetry, split into two chiral multiplets of opposite chirality,
$H=\left(H_L,H_R\right)$. At the fixed points, either $H_L$ or $H_R$ is projected out. For the
twisted sector fields, which come in three copies, the resulting spectrum contains two left- and one
right handed chiral multiplet or vice versa. Hence, two of these can always be combined to form a
singlet state $H_L H_R$ and be decoupled by an effective mass term
involving vevs of singlet fields (chosen such as to respect the string
selection rules).

At $n_2=0$, there are a number of non-singlet $\SU5$ multiplets
arising from the bulk $\SU6$ representations. Specifically, we obtain the following
left-chiral multiplets:
\begin{subequations}
\begin{align}
  \boldsymbol{35} &\longrightarrow \boldsymbol{5} +  \bar{\boldsymbol{5}}\\
  9\times\left(\boldsymbol{6} +  \bar{\boldsymbol{6}}\right)&\longrightarrow
  8\times\boldsymbol{5}+ 11\times\bar{\boldsymbol{5}}\\
  \boldsymbol{20} &\longrightarrow \boldsymbol{10} +  \bar{\boldsymbol{10}}
\end{align}
\end{subequations}
Among the $\boldsymbol{5}$-plets originating in the $\boldsymbol{6}$'s
of Eq.~(\ref{eq:t2t4}), six pairs can be decoupled immediately by
giving vevs to three non-Abelian singlets. Among the remaining $\boldsymbol{5}$-plets, two more
pairs can 
be decoupled in a second step, so that finally only 
two generations of $\bar{\boldsymbol{5}}\oplus\boldsymbol{10}$ and a
pair of $\boldsymbol{5} \oplus\bar{\boldsymbol{5}}$ Higgses
remain. The exotics located at $n_2=1$ can also be decoupled by
singlet vevs. 

\begin{figure}
  \centering
  \begin{picture}(230,200)(0,0)%\EBox(230,200)(0,00)
        \SetOffset(-55,-25)
        \put(35,15){\includegraphics[height=6cm]{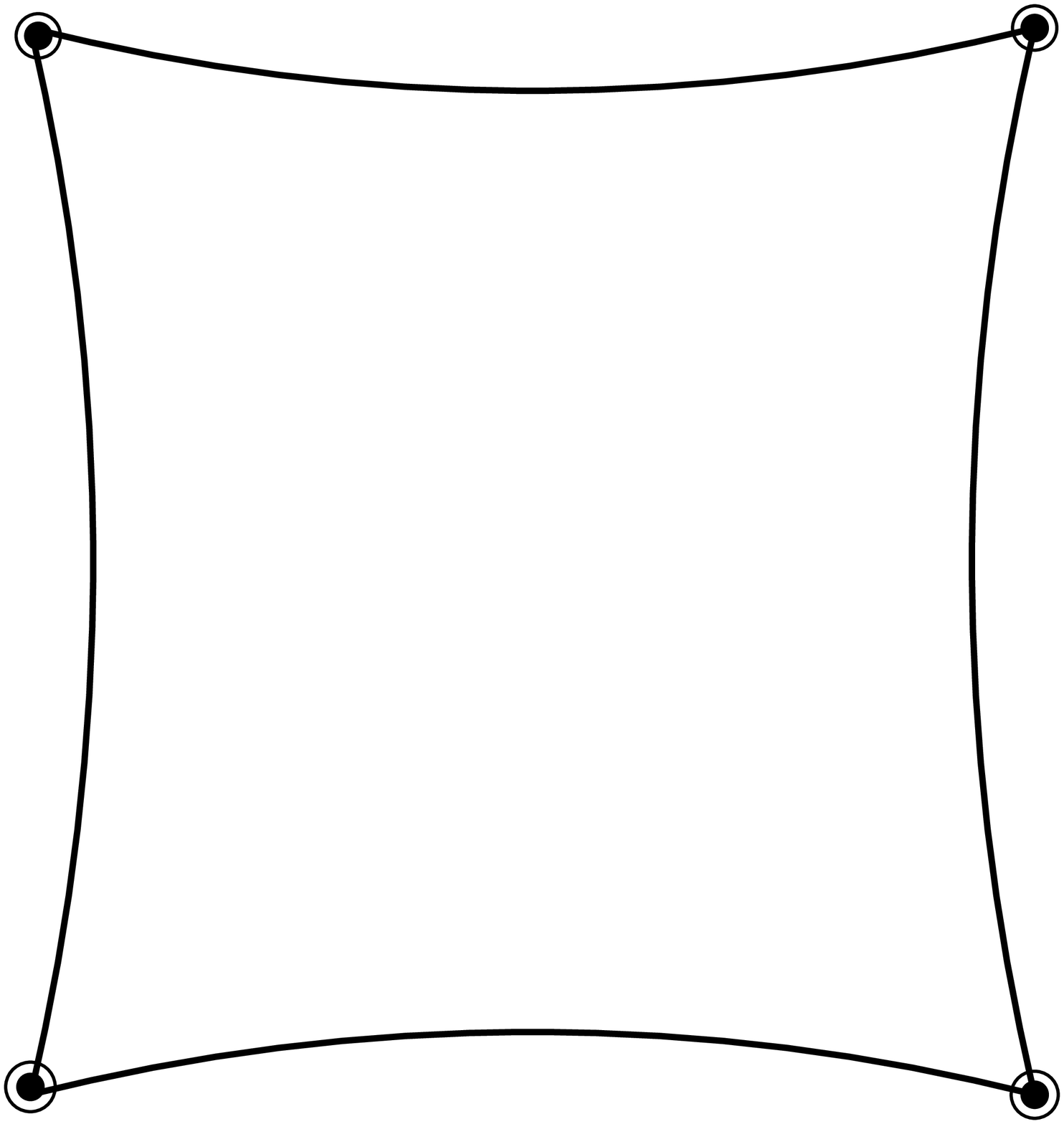}}
          \Text(215,141)[rb]{ {$\SU6  $}}
          \Text(70,212)[lb]{ {  $\SU5  $}}
          \Text(70,28)[lb]{ {  $\SU5  $}}
          \Text(210,212)[lb]{ {  $\SU2 \times \SU4 $}}
          \Text(210,28)[lb]{ {  $\SU2 \times \SU4  $}}
          \Text(195,132)[lb]{$\mathbf{35}$}
          \Text(180,119)[lb]{{$9 \times \left( {\bf 6} + {\bf \overline{6}} \right)$}}
          \Text(195,109)[lb]{{${\bf 20}$}}
          \Text(55,195)[lb]{  {  $ {\bf \overline{5}} +{\bf 10} $}}
          \Text(55,47)[lb]{  {  ${\bf \overline{5}} +{\bf 10}  $}}
%          \Text(245,195)[lb]{  {  \color{red} $4 \times {\bf 2}$ }}
%          \Text(245,47)[lb]{  {  \color{red} $4 \times {\bf 2}$ }}
          \LongArrow(178,136)(100,202)
          \COval(139,169)(17,27)(0){Black}{White}
          \Text(139,177)[]{$\boldsymbol{5}\oplus\bar{\boldsymbol{5}}$}
          \Text(139,166)[]{$2\times \bar{\mathbf{5}}\oplus\mathbf{10}$}
          \LongArrow(178,114)(100,48)
          \COval(139,81)(17,27)(0){Black}{White}
          \Text(139,89)[]{$\boldsymbol{5}\oplus\bar{\boldsymbol{5}}$}
          \Text(139,78)[]{$2\times \bar{\mathbf{5}}\oplus\mathbf{10}$}
      \end{picture}
  \caption{The $\mathbbm{Z}_2$ projection selects one pair of Higgses and two standard model
    generations at $n_2=0$, where $\SU5$ remains unbroken. At $n_2=1$, $\SU5$ is broken and only
    split multiplets survive, containing the Higgs doublets and one net generation. \label{fig:proj}}
\end{figure}

\subsection{Yukawa Couplings}
After decoupling, we have four standard model generations: Two in the bulk and one at each fixed point with
$n_2=0$ (see Fig.~\ref{fig:proj}). They can have a local superpotential with Yukawa couplings to the
Higgses,
\begin{align}
  W_\text{Yukawa}&= C_{ij}^{(u)} \boldsymbol{10}_i\boldsymbol{10}_j H_u + C_{ij}^{(d)}
  \bar{\boldsymbol{5}}_i\boldsymbol{10}_j H_d  \,,
\end{align}
where the coupling matrices $C_{ij}^{(u)}$ and $C_{ij}^{(d)}$ contain singlet vevs up to
$\mathcal{O}\!\left(8\right)$. The choice of Higgs fields is not unique: The model can have no,
partial or full gauge--Higgs unification, i.e.~one can choose the
Higgses to come from bulk $\boldsymbol{6}$'s or from the adjoint
$\boldsymbol{35}$. An attractive feature of identifying at least $H_u$
with the extra-dimensional component of the gauge field is that then
the top Yukawa coupling is given by the gauge coupling at hence is
naturally large.  

The $\mathbbm{Z}_2$ projection at the other fixed points finally breaks $\SU5$ and  projects out one
standard model generation from the two bulk 
$\bar{\boldsymbol{5}}\oplus\boldsymbol{10}$ fields -- the remaining fields again have the quantum
numbers of $\bar{\boldsymbol{5}}\oplus\boldsymbol{10}$, but the fact that they originate from
different split multiplets avoids the mostly unsuccessful $\SU5$ mass  relations. In the same way,
the $\mathbbm{Z}_2$ projection solves the doublet--triplet splitting problem by projecting the
unwanted triplets from the Higgs $\boldsymbol{5}$-plets.

\section{Vacuum Configurations}
To decouple unwanted states and generate Yukawa couplings, we need to give vevs to a number of
singlet fields. To respect supersymmetry, the $D$-term potential needs to vanish. For non-anomalous
symmetries, this can be accomplished by finding gauge invariant holomorphic monomials
$I=\prod\phi_i^{n_i}$ in the fields. Each of these monomials defines a $D$-flat direction by
\begin{align}
  \left<\phi^\dagger_i\right> =\frac{\partial I}{\partial \phi_i}\,.
\end{align}
The presence of the anomalous $U(1)$ additionally requires a monomial with negative anomalous
charge. This stabilises some of the flat directions at $\sim \xi$. Indeed, we find suitable
monomials such that the four-dimensional FI term is cancelled and the
$D$-term potential vanishes. It is intriguing at this point that the
scale of the FI terms is $\sim M_text{GUT}$.

\section{Summary}
We have constructed a local six-dimensional GUT from the heterotic string with the MSSM as its
low-energy limit. Two standard model generations are localised at fixed points with local $\SU5$
symmetry, while one is composed of split bulk multiplets. Also the Higgses are realised as split
multiplets, hence doublet--triplet splitting is automatic. We can find semirealistic vacuum
configurations respecting supersymmetry. As opposed to a direct four-dimensional limit, the
decoupling of exotic states is much more transparent due to the larger symmetry. 
The chosen vacuum is not phenomenologically viable: There is no $R$-parity forbidding proton decay,
and the electron and down quark are massless. However, this model is a member of a large class
(``mini-landscape'') of similar models \cite{lnx06,lnx07} where these problems have been addressed
and it seems likely that the phenomenology can be improved.  The main theoretical questions are
related to moduli stabilisation, the r\^ole of bulk field profiles and the blowup of singularities
which will be addressed elsewhere.

\subsection*{Acknowledgements}
I would like to thank my collaborators W.~Buchm\"{u}ller and J.~Schmidt.

\end{document}